\input harvmac

\def\Title#1#2{\rightline{#1}\ifx\answ\bigans\nopagenumbers\pageno0\vskip1in
\else\pageno1\vskip.8in\fi \centerline{\titlefont #2}\vskip .5in}
%

%%%%%%%%%%%%%%%%%%
%
% Figure macros, SBG 3/03
%
\ifx\includegraphics\UnDeFiNeD\message{(NO graphicx.tex, FIGURES WILL BE IGNORED)}
\def\figin#1{\vskip2in}% blank space instead
\else\message{(FIGURES WILL BE INCLUDED)}\def\figin#1{#1}
\fi
\def\Fig#1{Fig.~\the\figno\xdef#1{Fig.~\the\figno}\global\advance\figno
 by1}
%
%  Ifig   usage:
%
%         \Ifig{\Fig\figlabel}{caption}{figfile}{hsize}
%
% where vsize is the desired vertical size of the figure in truein
%

%
%defs
%
\font\ticp=cmcsc10

\def\hf{{1\over 2}}
\def\cnl{C^\lam_l}
\def\calo{{\cal O}}
\def\e{\varepsilon}
\def\lam{\lambda}
\def\q{{\bf q}}
\def\x{{\bf x}}
\def\eg{{\it e.g.}}
\def\roughly#1{\mathrel{\raise.3ex\hbox{$#1$\kern-.75em\lower1ex\hbox{$\sim$}}}}
%
%refs
%
\lref\Hawkunc{
  S.~W.~Hawking,
  ``Breakdown Of Predictability In Gravitational Collapse,''
  Phys.\ Rev.\ D {\bf 14}, 2460 (1976).
  %%CITATION = PHRVA,D14,2460;%%
}
\lref\ACVone{
  D.~Amati, M.~Ciafaloni and G.~Veneziano,
  ``Superstring Collisions at Planckian Energies,''
  Phys.\ Lett.\  B {\bf 197}, 81 (1987).
  %%CITATION = PHLTA,B197,81;%%
}
\lref\ACVtwo{
  D.~Amati, M.~Ciafaloni and G.~Veneziano,
  ``Classical and Quantum Gravity Effects from Planckian Energy Superstring
  Collisions,''
  Int.\ J.\ Mod.\ Phys.\  A {\bf 3}, 1615 (1988).
  %%CITATION = IMPAE,A3,1615;%%
}
\lref\KabatTB{
  D.~Kabat and M.~Ortiz,
  ``Eikonal Quantum Gravity And Planckian Scattering,''
  Nucl.\ Phys.\  B {\bf 388}, 570 (1992)
  [arXiv:hep-th/9203082].
  %%CITATION = NUPHA,B388,570;%%
}
\lref\LQGST{
  S.~B.~Giddings,
  ``Locality in quantum gravity and string theory,''
  Phys.\ Rev.\  D {\bf 74}, 106006 (2006)
  [arXiv:hep-th/0604072].
  %%CITATION = PHRVA,D74,106006;%%
}
\lref\AmatiXE{
  D.~Amati, M.~Ciafaloni and G.~Veneziano,
  ``Higher Order Gravitational Deflection And Soft Bremsstrahlung In Planckian
  Energy Superstring Collisions,''
  Nucl.\ Phys.\  B {\bf 347}, 550 (1990).
  %%CITATION = NUPHA,B347,550;%%
}
\lref\GiddingsHH{
  S.~B.~Giddings,
  ``Black holes and massive remnants,''
  Phys.\ Rev.\  D {\bf 46}, 1347 (1992)
  [arXiv:hep-th/9203059].
  %%CITATION = PHRVA,D46,1347;%%
}
\lref\tHooholo{
  G.~'t Hooft,
  ``Dimensional reduction in quantum gravity,''
  arXiv:gr-qc/9310026.
  %%CITATION = GR-QC 9310026;%%
  }
\lref\Sussholo{
  L.~Susskind,
  ``The World as a hologram,''
  J.\ Math.\ Phys.\  {\bf 36}, 6377 (1995)
  [arXiv:hep-th/9409089].
  %%CITATION = HEP-TH 9409089;%%
}
\lref\AiSe{
  P.~C.~Aichelburg and R.~U.~Sexl,
  ``On the Gravitational field of a massless particle,''
  Gen.\ Rel.\ Grav.\  {\bf 2}, 303 (1971).
  %%CITATION = GRGVA,2,303;%%
}
\lref\BPS{
  T.~Banks, L.~Susskind and M.~E.~Peskin,
  ``Difficulties For The Evolution Of Pure States Into Mixed States,''
  Nucl.\ Phys.\  B {\bf 244}, 125 (1984).
  %%CITATION = NUPHA,B244,125;%%
}
\lref\MSpur{
  M.~Srednicki,
  ``Is purity eternal?,''
  Nucl.\ Phys.\  B {\bf 410}, 143 (1993)
  [arXiv:hep-th/9206056].
  %%CITATION = NUPHA,B410,143;%%
}
\lref\EaGi{
  D.~M.~Eardley and S.~B.~Giddings,
  ``Classical black hole production in high-energy collisions,''
  Phys.\ Rev.\  D {\bf 66}, 044011 (2002)
  [arXiv:gr-qc/0201034].
  %%CITATION = PHRVA,D66,044011;%%
}
\lref\QBHB{
  S.~B.~Giddings,
  ``Quantization in black hole backgrounds,''
  Phys.\ Rev.\  D {\bf 76}, 064027 (2007)
  [arXiv:hep-th/0703116].
  %%CITATION = PHRVA,D76,064027;%%
}
\lref\Astrorev{
  A.~Strominger,
  ``Les Houches lectures on black holes,''
  arXiv:hep-th/9501071.
  %%CITATION = HEP-TH 9501071;%%
}
\lref\GMH{
  S.~B.~Giddings, D.~Marolf and J.~B.~Hartle,
  ``Observables in effective gravity,''
  Phys.\ Rev.\  D {\bf 74}, 064018 (2006)
  [arXiv:hep-th/0512200].
  %%CITATION = PHRVA,D74,064018;%%
}
\lref\GiddingsNU{
  S.~B.~Giddings and D.~Marolf,
  ``A global picture of quantum de Sitter space,''
  Phys.\ Rev.\  D {\bf 76}, 064023 (2007)
  [arXiv:0705.1178 [hep-th]].
  %%CITATION = PHRVA,D76,064023;%%
}
\lref\GaGi{
  M.~Gary and S.~B.~Giddings,
  ``Relational observables in 2d quantum gravity,''
  arXiv:hep-th/0612191, Phys.\ Rev.\  D {\bf 75} 104007 (2007).
  %%CITATION = HEP-TH/0612191;%%
}
\lref\YoNa{
  H.~Yoshino and Y.~Nambu,
  ``Black hole formation in the grazing collision of high-energy particles,''
  Phys.\ Rev.\  D {\bf 67}, 024009 (2003)
  [arXiv:gr-qc/0209003].
  %%CITATION = PHRVA,D67,024009;%%
}
\lref\Froiss{
  M.~Froissart,
  ``Asymptotic behavior and subtractions in the Mandelstam representation,''
  Phys.\ Rev.\  {\bf 123}, 1053 (1961).
  %%CITATION = PHRVA,123,1053;%%
}
\lref\CeMa{F. Cerulus and A. Martin, ``A lower bound for large angle elastic 
scattering at high energies,''  Phys. Lett. {\bf 8}, 80 (1964).}
\lref\Mart{A. Martin, ``Minimal interactions at very high transfers,''
 Nuov. Cim. {\bf 37}, 671 (1965).}
\lref\WeinbergNX{
  S.~Weinberg,
  ``Infrared photons and gravitons,''
  Phys.\ Rev.\  {\bf 140}, B516 (1965).
  %%CITATION = PHRVA,140,B516;%%
}
\lref\Szego{G. Szego, {\sl Orthogonal polynomials,} (1939. American mathematical society.  Colloquium Publications.)}
\lref\FPVV{
  M.~Fabbrichesi, R.~Pettorino, G.~Veneziano and G.~A.~Vilkovisky,
  ``Planckian energy scattering and surface terms in the gravitational
 action,''
  Nucl.\ Phys.\  B {\bf 419}, 147 (1994).
  %%CITATION = NUPHA,B419,147;%%
}
\lref\GiddingsBE{
  S.~B.~Giddings,
 ``(Non)perturbative gravity, nonlocality, and nice slices,''
  Phys.\ Rev.\  D {\bf 74}, 106009 (2006)
  [arXiv:hep-th/0606146].
  %%CITATION = PHRVA,D74,106009;%%
}
\lref\VenezianoER{
  G.~Veneziano,
  ``String-theoretic unitary S-matrix at the threshold of black-hole
  production,''
  JHEP {\bf 0411}, 001 (2004)
  [arXiv:hep-th/0410166].
  %%CITATION = JHEPA,0411,001;%%
}
\lref\Hsu{
  S.~D.~H.~Hsu,
 ``Quantum production of black holes,''
  Phys.\ Lett.\  B {\bf 555}, 92 (2003)
  [arXiv:hep-ph/0203154].
  %%CITATION = PHLTA,B555,92;%%
}
\lref\GiddingsSJ{
  S.~B.~Giddings,
  ``Black hole information, unitarity, and nonlocality,''
  Phys.\ Rev.\  D {\bf 74}, 106005 (2006)
  [arXiv:hep-th/0605196].
  %%CITATION = PHRVA,D74,106005;%%
}
\lref\ArkaniHamedKY{
  N.~Arkani-Hamed, S.~Dubovsky, A.~Nicolis, E.~Trincherini and G.~Villadoro,
  ``A Measure of de Sitter Entropy and Eternal Inflation,''
  arXiv:0704.1814 [hep-th].
  %%CITATION = ARXIV:0704.1814;%%
}
\lref\AmatiTB{
  D.~Amati, M.~Ciafaloni and G.~Veneziano,
  ``Effective action and all order gravitational eikonal at Planckian
  energies,''
  Nucl.\ Phys.\  B {\bf 403}, 707 (1993).
  %%CITATION = NUPHA,B403,707;%%
}
\lref\Sold{
  M.~Soldate,
  ``Partial Wave Unitarity and Closed String Amplitudes,''
  Phys.\ Lett.\  B {\bf 186}, 321 (1987).
  %%CITATION = PHLTA,B186,321;%%
}
\lref\MuSo{
  I.~J.~Muzinich and M.~Soldate,
  ``High-Energy Unitarity of Gravitation and Strings,''
  Phys.\ Rev.\  D {\bf 37}, 359 (1988).
  %%CITATION = PHRVA,D37,359;%%
}
\lref\BaFi{
  T.~Banks and W.~Fischler,
  ``A model for high energy scattering in quantum gravity,''
  arXiv:hep-th/9906038.
  %%CITATION = HEP-TH/9906038;%%
}
\lref\CFV{
  M.~Chaichian, J.~Fischer and Yu.~S.~Vernov,
  ``Generalization Of The Froissart-Martin Bounds To Scattering In A Space-Time
  Of General Dimension,''
  Nucl.\ Phys.\  B {\bf 383}, 151 (1992).
  %%CITATION = NUPHA,B383,151;%%
}
\lref\GGM{
  S.~B.~Giddings, D.~J.~Gross and A.~Maharana,
  ``Gravitational effects in ultrahigh-energy string scattering,''
  arXiv:0705.1816 [hep-th], to appear in Phys. Rev. D.
  %%CITATION = ARXIV:0705.1816;%%
}
\lref\UQM{ S.~B.~Giddings,
  ``Universal quantum mechanics,''
  arXiv:0711.0757 [quant-ph].
  %%CITATION = ARXIV:0711.0757;%%
}
\lref\MyersUN{
  R.~C.~Myers and M.~J.~Perry,
  ``Black Holes In Higher Dimensional Space-Times,''
  Annals Phys.\  {\bf 172}, 304 (1986).
  %%CITATION = APNYA,172,304;%%
}
\lref\GiddingsPJ{
  S.~B.~Giddings,
  ``Black holes, information, and locality,''
  arXiv:0705.2197 [hep-th].
  %%CITATION = ARXIV:0705.2197;%%
}
\lref\SGinfo{S.~B.~Giddings,
  ``Quantum mechanics of black holes,''
  arXiv:hep-th/9412138\semi
  %%CITATION = HEP-TH 9412138;%%
  ``The black hole information paradox,''
  arXiv:hep-th/9508151.
  %%CITATION = HEP-TH 9508151;%%
}
\lref\WABHIP{
  S.~B.~Giddings,
  ``Why Aren't Black Holes Infinitely Produced?,''
  Phys.\ Rev.\  D {\bf 51}, 6860 (1995)
  [arXiv:hep-th/9412159].
  %%CITATION = PHRVA,D51,6860;%%
}
\Title{\vbox{\baselineskip12pt
}}
{\vbox{\centerline{High-energy gravitational scattering  }\centerline{ and black hole resonances}
}}
\centerline{{\ticp Steven B. Giddings\footnote{$^\ast$}{Email address: giddings@physics.ucsb.edu} and Mark Srednicki\footnote{$^\dagger$}{Email address: mark@physics.ucsb.edu} } }
\centerline{\sl Department of Physics}
\centerline{\sl University of California}
\centerline{\sl Santa Barbara, CA 93106}
\bigskip
\centerline{\bf Abstract}
Aspects of super-planckian gravitational scattering and black hole formation are investigated, largely via a partial-wave representation.  At large and decreasing impact parameters, amplitudes are expected to be governed by single graviton exchange, and then by eikonalized graviton exchange, for which partial-wave amplitudes are derived.  In the near-Schwarzschild regime, perturbation theory fails.  However, general features of gravitational scattering associated with black hole formation suggest a particular form for amplitudes, which we express as a {\it black hole ansatz}.  We explore features of this ansatz, including its locality properties.  These  amplitudes satisfy neither the Froissart bound, nor apparently  the more fundamental property of polynomial boundedness, through which locality is often encoded in an S-matrix framework.  Nevertheless, these amplitudes do satisfy a macroscopic form of causality, expressed as a polynomial bound for the forward-scattering amplitude.

%\draftmode
\Date{}

\newsec{Introduction}

The puzzles of quantum gravity come into particularly sharp focus in non-perturbative contexts, such as black holes and quantum cosmology.  One avenue towards better framing these puzzles and investigating their possible resolution is via study of high-energy gravitational scattering.  Above the Planck energy, such scattering can probe the non-perturbative sector, through what is classically described as black hole formation.  However, unlike in cosmology, in the scattering context one can place such questions in a more  tractable framework because one expects a simple description of asymptotic in and out states.  Thus, working about a flat background geometry, one can investigate properties of the gravitational S-matrix.  

The description of such scattering hinges on the fate of quantum black holes.  For example, if Hawking's original picture of information loss\refs{\Hawkunc} were correct, scattering should instead be described by a superscattering matrix, that parametrizes non-unitary evolution of density matrices.  Likewise, a scenario with black hole remnants would have important consequences for final state properties.  However, both of these proposals appear to lead to unacceptable physics (violations of energy conservation\BPS\ and/or Lorentz invariance\MSpur, and instability,\foot{See \eg\ \WABHIP.} respectively), and thus the belief has grown that the resolution of the information paradox\foot{For reviews, see \refs{\SGinfo,\Astrorev}.}  will involve unitary evolution without remnants.  Such a scenario, in which information escapes in Hawking radiation from a macroscopic black hole,  appears to require macroscopic violations of locality;\foot{For a short summary of some issues in locality of gravity, see \GiddingsPJ.} early proposals in this direction include \GiddingsHH\ and the holographic ideas of 't Hooft\tHooholo\ and Susskind\Sussholo.

While the question of the exact mechanism for unitarity restoration in black hole evaporation remains mysterious, we consider it a likely result.  Moreover, in an S-matrix context, one doesn't necessarily have to commit to such an internal explanation in order to investigate some aspects of the physics; assuming that the physics is indeed unitary, one can explore properties of the corresponding S-matrix directly.  

If there is indeed a fundamental breakdown of locality in nonperturbative gravity, it is ultimately important to characterize this breakdown and understand its consequences.  Even formulating the principle of locality is difficult in a gravitational theory.  Due to diffeomorphism invariance, precise local observables appear not to exist, and instead seem to be recovered only approximately in certain states from ``proto-local" observables, as is described in \refs{\GMH,\GaGi}.  This complicates usual formulations of microcausality phrased in terms of commutativity of local operators.  But another set of criteria for locality involve properties of the S-matrix, particularly various bounds on its high-energy behavior.  A basic approach of this paper will be to assume expected general properties of high-energy scattering, such as black hole formation and unitary evolution; the result is a {\it black hole ansatz} for the structure of partial-wave scattering amplitudes.  One can then inquire whether this ansatz yields an S-matrix that respects usual locality criteria.  We will find that it apparently does not -- not only does it violate the Froissart bound\Froiss, but also it does not respect a more fundamental constraint of locality, polynomial boundedness of amplitudes.  Interestingly and importantly, though, it does appear to respect constraints of causality through a polynomial bound on the forward scattering amplitude.  Such scattering behavior seems like a potentially important further clue about the status of locality in gravity, and about the ultimate structure of the quantum theory.

In outline, in section two we will discuss some basic issues of scattering in gravity, and in particular argue that an S-matrix approach is plausibly justifiable in gravity.  We also summarize some of the important regimes for gravitational scattering, organized by decreasing impact parameter, and review aspects of partial-wave decompositions in general dimension.  Section three then treats scattering in the large impact parameter regime; at the longest distances this is simply Born exchange, and at shorter distances the Born amplitudes are unitarized by the eikonal amplitudes.  The latter correspond nicely to a classical description, providing additional evidence that one is justified in relying on features of the semiclassical picture in the strong-gravity regime, where the impact parameter reaches the Schwarzschild radius given by the center-of-mass energy.  Section four turns to description of the quantum physics in this regime, in terms of very general assumptions about properties of black holes that would appear as resonances in the scattering amplitude.  As a result, we outline a {\it black hole ansatz} for the two-two exclusive S-matrix, given in terms of its partial-wave amplitudes.  Section five then investigates asymptotic properties of this ansatz, and in particular the status of the Froissart\Froiss\ and Cerulus-Martin\CeMa\ bounds, and polynomial boundedness, revealing the apparently nonlocal but causal behavior.  

The reader wishing to proceed directly to the interesting features of the strong gravity regime may choose to quickly peruse section two, then read sections four and five.

\newsec{Basics of gravitational scattering}

\subsec{Framework}

Our interest is  gravitational scattering at energies well above the Planck scale.  
We currently lack a complete description of quantum-gravitational dynamics making predictions for such scattering.  Moreover, there are significant indications that such a theory will not simply be a local field theory, say based on quantization of the Einstein action supplemented by some matter terms.\foot{For some further discussion, see \refs{\GiddingsPJ}.}  For that reason, we will fall back to a more basic viewpoint, and inquire about the essential features of the S-matrix describing this scattering.  

Given an underlying microphysics, one ordinarily expects to be able to compute the S-matrix (in cases where it exists); conversely, given the S-matrix, one also expects to be able to learn a great deal about the microphysics.  Our specific approach will be to investigate properties of the S-matrix that follow from generic behavior expected of a gravitational theory.  In the light of the fact that gravity is likely  not described by a local field theory, it is important to outline what we believe are valid assumptions about such a theory.

Specifically, we will assume that the theory is quantum mechanical.\foot{A proposal for a quantum framework sufficiently general to incorporate quantum gravity appears in \refs{\UQM}.}  We will moreover assume that states of the theory exist corresponding to excitations of flat Minkowski space, and can be labeled by  ``in" and ``out" basis representations as with familiar S-matrix theory.  These states include a vacuum, and excitations about this described as asymptotic multiparticle states of widely separated particles.  These should include graviton states, as well as states of the matter fields.  (In the case of string theory, these asymptotic states are multi-string states of the stable string modes.)  We will also assume that the theory is Lorentz invariant.  By this we mean that there is a unitary action of the Lorentz group on the asymptotic states, such that the vacuum is invariant, and such that the S-matrix (see below) is Lorentz covariant.  However, it is important to stress that we will {\it not} necessarily assume that there is a more local notion of Lorentz invariance, or even a precise local notion of space and time.  But we will assume that there is a regime where there is a  semiclassical approximation to the full quantum dynamics  described by general relativity plus matter fields.  

As a simple example, one might consider a theory that in the semiclassical limit corresponds to gravity minimally coupled to a single 
massive 
scalar field.  This theory will be kept in mind as a model for generic dynamics, although one may wish to consider more elaborate matter contents, or strings.

Given a labeling of asymptotic states $|\alpha\rangle_{in}$, $|\alpha\rangle_{out}$,
as described above,  one may define the S-matrix,
\eqn\Sdef{S_{\alpha\beta} = {}_{out}\langle\alpha|\beta\rangle_{in}\ .}
However, due to masslessness of the graviton (and possibly other fields) there may be subtleties in its definition resulting from soft particle emission and corresponding infrared divergences.  While our ultimate interest is in four non-compact dimensions, we will largely sidestep these issues by working in spacetime dimension $D>4$.  (We thus implicitly assume these issues are not fundamental.)  Dimension $D\geq5$ is sufficient to eliminate soft-graviton divergences, and $D\geq7$ is needed for existence of the total cross section. Even here, as we will see, masslessness of the graviton has various consequences for analyticity properties of the S-matrix.

\subsec{Scattering regimes}

We will particularly focus on  scattering of massive particles, \eg\ scalars, of momenta $p_1$, $p_2$.  
The Mandelstam parameters are
\eqn\sdef{s=-(p_1+p_2)^2 = E^2\ ,} 
and, in the case of exclusive $2\rightarrow2$ scattering to particles with momenta $p_3$, $p_4$,
\eqn\tudef{t=-(p_1-p_3)^2\ ,\ u = (p_1-p_4)^2\ .}
As is outlined in \refs{\LQGST,\GGM}, there are different regimes with different dynamics dominating the scattering behavior; much earlier work on this scattering has been done by Amati, Ciafaloni, and Veneziano\refs{\ACVone\ACVtwo\AmatiXE-\AmatiTB}.  These can be classified in terms of $s$ and $t$, or even more intuitively, in terms of the center-of-mass (CM) energy $E$ and impact parameter $b$, and are determined by the $D$-dimensional gravitational constant $G_D$, as well as other parameters such as for example the string scale, etc.  The three regions of generic interest are:
\item{1.} The Coulomb (or Born) regime.  Here the scattering is well-described by one-graviton exchange.  This regime corresponds to  $b\roughly> (G_D E^2)^{1/ (D-4)}$.  
\item{2.} The eikonal regime.  Here scattering is well-described by the sum of ladder diagrams, exponentiating the single-graviton exchange diagram.  This closely corresponds to classical gravitational scattering, and is expected to be valid in a region bounded by
$(G_D E^2)^{1/ (D-4)}\roughly> b\roughly> R_S(E) \sim (G_D E)^{1/(D-3)}$.
\item{3.} The strong gravity or black hole regime.  Here a perturbative description of the dynamics fails.\foot{Even apparently in string theory.}  In a semiclassical picture this regime corresponds to black hole formation; the appropriate quantum description of this regime is a central problem.  This occurs at impact parameters $R_S(E)\roughly> b$.

\noindent String theory, or other theories representing new dynamics ({\it e.g.} composite structures) at a given scale,  add possible subregimes, where tidal excitation of strings or composite structure, {\it etc.}, can play a role in the description of the asymptotics.  

We will also work in an angular-momentum (partial-wave) representation.  While to be precise, one should convert to this representation via an impact parameter-angular momentum transformation, 
the above regimes can be approximately converted into regimes for $l$ using the classical relation 
\eqn\lvsb{l=Eb/2\ .}
In particular, this leads to the definition of a critical angular momentum,
\eqn\Ldef{L(E)=ER_S(E)/2\propto \left(G_DE^{D-2}\right)^{1/(D-3)}\ ,}
below which one enters the strong-gravitational regime.

\subsec{Partial wave essentials}

A partial-wave representation will be particularly useful for describing features of the scattering.  
This subsection will collect some of the basic formulas needed for our analysis, and more are provided in the appendix.
The transition matrix element $T$ for exclusive scattering may be defined via $S=1+i{\cal T}$, with
\eqn\tmat{{\cal T}_{p_3p_4,p_1p_2} = (2\pi)^D \delta^D(p_1+p_2-p_3-p_4) T(s,t)\ .}
This has partial-wave expansion\Sold
\eqn\Tsimp{T(s,t) = \psi_\lam s^{2-D/2} \sum_{l=0}^\infty (l+\lam) C_l^\lam(\cos \theta) f_l(s)\ }
where $\lam=(D-3)/2$,  
\eqn\psidef{\psi_\lam = 2^{4\lam+3}\pi^\lam \Gamma(\lam)\ ,}
and $C_l^\lam$ are Gegenbauer polynomials.  Here we use the ultrarelativistic relation for the scattering angle,
\eqn\thetadeff{-t/s = q^2/s=\sin^2{\theta\over 2}\ .}

The individual partial-wave amplitudes $f_l(s)$ can be derived from \Tsimp\ using the orthogonality of the $C^\lam_l$ (see the appendix) with the result
\eqn\pwderiv{ f_l(s) = {s^{(D-4)/2}\over \gamma_D \cnl(1)} \int_0^\pi d\theta\, \sin^{D-3}\theta\, \cnl(\cos\theta) T(s,t)\ ,}
with 
\eqn\gammaDef{\gamma_D = 2 \Gamma\left({D-2\over 2}\right) (16\pi)^{(D-2)/2}\ .}
The partial-wave amplitudes should satisfy the unitarity conditions\refs{\Sold}
\eqn\pwunit{{\rm Im} f_l(s)\geq |f_l(s)|^2\ .}
The general such amplitudes can be written in terms of the real parameters $\delta_l$ and $\beta_l$:
\eqn\pwsoln{ f_l(s) = {e^{2i\delta_l(s) - 2 \beta_l(s)}-1\over 2i}\ .}
The real and imaginary parts, and the norm, are
\eqn\reaf{r_l(s) = {\rm Re} f_l(s) = \hf e^{-2\beta_l(s)} \sin(2\delta_l(s))\ ,}
\eqn\imf{a_l(s) = {\rm Im} f_l(s) = \hf\left[ 1-e^{-2\beta_l(s)} \cos(2\delta_l(s))\right]\ ,}
\eqn\normof{|f_l(s)|^2 = {1\over 4}\left[ 1- 2 e^{-2\beta_l(s)}  \cos(2\delta_l(s))+ e^{-4 \beta_l(s)}\right]\ .}

\newsec{Scattering in the eikonal and Born regimes}

In order to begin understanding the partial-wave description of gravitational scattering, we first investigate this scattering in the large impact parameter regime.  As stated, in this regime ladder and crossed-ladder diagrams are dominant contributions to scattering.  Working in the approximation $-t/s\ll 1$, one can sum all such amplitudes to obtain the {\it eikonal amplitude}, 
\refs{\ACVone,\ACVtwo,\MuSo,\KabatTB}
\eqn\Teik{ iT_{\rm eik}(s,t) = 2s \int d^{D-2} x_\perp e^{-i\q_\perp \cdot \x_\perp}(e^{i\chi(x_\perp,s)} -1)\ .}
Here $\q_\perp$ is the $(D-2)$-vector component of the momentum transfer $q$ perpindicular to the CM collision axis; in terms of CM variables, $|\q_\perp| = (\sqrt{s} \sin\theta)/2$.  (Note that at small angles, $q_\perp^2\approx q^2$.)   
$\chi(x_\perp,s)$ is the eikonal phase, which  is given by $1/s$ times
the Fourier transform of the one-graviton exchange or tree amplitude, 
\eqn\Ttree{T_{\rm tree}(s,t)=-8\pi G_D  s^2/t\ .}
This Fourier transform is taken with respect to the transverse momentum variable, giving
\eqn\eikphase{\eqalign{\chi(x_\perp,s) &= {1\over 2s}\int{d^{D-2}k_\perp\over(2\pi)^{D-2}}\,
e^{-i{\bf k}_\perp\cdot\x_\perp}T_{\rm tree}(s,-k^2_\perp) \cr
&= {4\pi\over (D-4) \Omega_{D-3}} {G_D s\over x_\perp^{D-4}}\ ,}}
%&= \pi^{-(D-4)/2}\Gamma({\textstyle{D-4\over2}}){G_D s\over x_\perp^{D-4}}\ .}}
%
where 
\eqn\vols{ \Omega_n={2\pi^{(n+1)/2}\over \Gamma[(n+1)/2]}}
is the volume of the unit $n$-sphere.

The small expansion parameter justifying use of the leading eikonal amplitude \Teik\  is thus $\theta\ll 1$. Higher-order loop diagrams corresponding to connecting the two external lines with multipoint graviton tree diagrams can be seen\refs{\ACVtwo} to be
subleading in an expansion in $(R_S/b)^{D-3}\sim \theta$, and thus are also small for $\theta\ll 1$.  

Partial-wave amplitudes in this regime are straightforwardly derived.  Since $\chi(x_\perp,s)$ is a function only of the magnitude 
of the $(D-2)$-vector $\x_\perp$,
we can integrate over angles in eq.~\Teik\ to get\refs{\MuSo}
\eqn\Teikrad{ T_{\rm eik}(s,t) = -2is(2\pi)^{(D-2)/2}q_\perp^{-(D-4)/2}
\int_0^\infty dx_\perp x_\perp^{(D-2)/2}J_{(D-4)/2}(q_\perp x_\perp)(e^{i\chi(x_\perp,s)} -1)\ ,}
where $J_\nu$ is a Bessel function.
We can now plug eq.~\Teikrad\ into eq.~\pwderiv\ to get the eikonal approximation for 
the partial-wave amplitudes,  
\eqn\fleik{ f^{\rm eik}_l(s) = 
-i^{l+1}{(4\pi)^{\lam+1}\over \gamma_D} 
{C^\lam_l(0)\over C_l^\lam(1)}s^{(\lam+1)/2}
\int_0^\infty dx_\perp\,x_\perp^\lam J_{l+\lam}(\half\sqrt{s}x_\perp)(e^{i\chi(x_\perp,s)} -1)\ ,}
where again $\lam = (D-3)/2$.  Note that the result is proportional to $C^\lam_l(0)$, which
vanishes for odd $l$; see Appendix A.  
This would follow from $t \leftrightarrow u$ crossing symmetry, but
we have not assumed that our initial particles are identical, and so should not expect
$t \leftrightarrow u$ crossing symmetry to hold as a general result.
Instead, it is an artifact of the small-angle approximation used to derive the  amplitude.  Such a small angle approximation cannot accurately predict fine-scale
structure of the partial-wave coefficients as a function of $l$, but should only be trusted
to give an overall envelope function that changes slowly with $l$.
We will revisit this issue below.

Defining a new integration variable $v= x_\perp\sqrt{s}/2l$ and plugging in the values of 
$C^\lam_l(0)$ and $C^\lam_l(1)$ from Appendix A, we get (for $l$ even)
\eqn\fleika{ f^{\rm eik}_l(s) 
= -i\,{2^{-\lam}\Gamma[\half(l+1)]l^{\lam+1}\over \Gamma[\half(l+1)+\lam]}
\int_0^\infty dv\,v^\lam J_{l+\lam}(lv)[e^{il\varepsilon(l,s)/v^{D-4}} -1]\ ,}
where
\eqn\chit{\eqalign{\varepsilon(l,s) &= (4\pi)^{-(D-4)/2}\Gamma({\textstyle{D-4\over2}})
{G_D s^{(D-2)/2}\over l^{D-3} }\cr
&= {\sqrt{\pi} (D-2) \Gamma[(D-4)/2]\over 4 \Gamma[(D-1)/2] }\left[{L(E)\over l}\right]^{D-3}\ .\cr}}
Here  $L(E)$ is the critical angular momentum defined in
eq.~\Ldef, with $E=s^{1/2}$.  We will be in the eikonal regime for $l \gg L(E)$; here $\varepsilon$ is small.
Since we are interested in high energies, we have $L(E)\gg 1$, and so we can
also take $l\gg1$ in eq.~\fleika.  We can then write the Bessel function as
\eqn\Jdef{ J_{l+\lam}(lv)={1\over 2\pi}\int_{-\pi}^{+\pi}d\phi\;e^{i(l+\lam)\phi-ilv\sin\phi}
+{\cal O}(1/l)\ ;}
this formula is exact if $l+\lam$ is an integer.  
Inserting this in eq.~\fleika, and taking the large-$l$ limit in the prefactor, we find
\eqn\fleikb{ f^{\rm eik}_l(s) 
= {-il\over 2\pi}\int_{-\pi}^{+\pi}d\phi
\int_0^\infty dv\,(ve^{i\phi})^\lam\; 
e^{il(\phi-v\sin\phi)}(e^{il\e/v^{D-4}} -1)+{\cal O}(1/l)\ ,}
We can now evaluate this double integral by stationary
phase.  Then we must minimize $S = \phi-v\sin\phi+\e/v^{D-4}$
with respect to both $\phi$ and $v$.  For $\e\ll 1$, the solution is $v=1+O(\e^2)$ and
$\phi=-\e+O(\e^3)$; at this point, $S=\e+O(\e^3)$, and the determinant of the matrix of the second derivatives of $S$ is $-1+O(\e^2)$.  We can then account for the $-1$ term in the integrand
by subtracting the result with $\e$ set to zero.
(A more careful analysis shows that this is correct up to corrections that are suppressed
by a relative factor of $\e^{1/2}$.)  For $l\gg1$ and $\e\ll1$, we thus find 
\eqn\fleikc{  f^{\rm eik}_l(s) 
= {-il\over 2\pi}\left({2\pi\over l}\right)(e^{il\e} -1)
=-i(e^{il\e} -1)}
for even $l$, and (as previously noted) zero for odd $l$.  Comparing with eq.~\pwsoln, we see
that this even-$l$ result is too large by a factor of two to be unitary.  However, as previously
discussed, we cannot trust the eikonal approximation to get the correct fine structure of the 
partial-wave amplitudes as a function of $l$, but only to give an envelope function that changes
slowly with $l$.  Since the odd-$l$ amplitudes are zero, we get this envelope function by taking
half of eq.~\fleikc, and applying it for both even and odd $l$.  This gives a unitary result,
with  the real and imaginary parts of the partial-wave phase shifts given by
\eqn\dbeik{ \eqalign{
\delta^{\rm eik}_l(s) &= {\sqrt{\pi} (D-2) \Gamma[(D-4)/2]\over 8 \Gamma[(D-1)/2] }{L(E)^{D-3}\over l^{D-4}}\ ,\cr
\beta^{\rm eik}_l(s)&=0\ .\cr}}
In the Coulomb regime, $\delta^{\rm eik}_l(s)\ll 1$, this is simply the partial-wave phase shift
corresponding to the tree-level Coulomb amplitude.  Thus we see that the eikonal approximation
provides a unitarization of the tree-level phase shifts.  

As one leaves the regime $\theta\ll 1$, corrections to the leading eikonal phase shifts \dbeik\ become important.  The leading contributions from the iterated tree graph exchanges have been argued\refs{\AmatiXE,\AmatiTB} to yield corrections to the eikonal phase \eikphase\ that correspond to higher-order classical corrections to the linearized metric.  Specifically, these corrections appear to match the classical picture, which is scattering of Aichelburg-Sexl shock waves\AiSe.  This serves as motivation\foot{Additional motivation is discussed in \refs{\BaFi,\Hsu}.} to rely on features of the semiclassical picture in the regime $b\sim R_S$.

One should also note that various corrections can lead to non-zero absorptive coefficients $\beta_l$.  A universal contribution comes from gravitational bremsstrahlung of soft gravitons.  
Using methods of \WeinbergNX, one can estimate their contributions to have parametric dependence
\eqn\softg{\beta_l^{SG}\sim {L^{3D-9}/ l^{3D-10}}\ }
in the eikonal regime.

There are also model-dependent contributions arising from the substructure of the scattered states.  An example of this  is scattering of strings: at sufficiently small impact parameter, tidal forces become large enough to excite internal vibrations of the scattered strings\LQGST.  If one scatters any other composite objects (for example, hydrogen atoms or neutrons), one likewise expects a tidal excitation threshold where the internal structure becomes excited.  In the string case, this excitation is the diffractive excitation found in \refs{\ACVone,\ACVtwo}, and the corresponding absorptive coefficients have size
\eqn\betats{\beta_l^{TS} \sim {(l_s E)^2\over L} \left({L\over l}\right)^{D-2}\ ,}
where $l_s$ is the string length.
Absorptive coefficients in other cases clearly depend on the details of the composite structure.

\newsec{The black hole ansatz}

\subsec{Quantum scattering}

As we have discussed, some basic features of high-energy gravitational scattering at large impact parameter are well-described by semiclassical and/or perturbative physics.  In particular, the leading eikonal diagrams represent the leading contribution to what is essentially classical scattering, working to lowest order in an expansion in $R_S/b$, and \refs{\AmatiXE,\AmatiTB} have argued that subleading eikonal diagrams correspond to higher-order terms in the expansion of the classical metric in $R_S/b$.  Thus, one has what appears to be a relatively reliable picture of the scattering for $b\roughly>R_S$.  However, such an expansion should clearly break down in the impact parameter regime $b\roughly< R_S$, where the perturbation theory apparently diverges.

This is the regime where the collision would classically form a black hole\refs{\EaGi,\YoNa}.  An explicit derivation of quantum amplitudes in this regime apparently requires knowledge of the non-perturbative dynamics of quantum gravity.  However, since in the classical picture the resulting black hole is large, and has weak curvature at the horizon, one might expect that at least crude features of the quantum dynamics are captured by the semiclassical  description.  
Semiclassically, one predicts black hole formation, and subsequent Hawking evaporation.  We will assume that this describes gross features of the true quantum dynamics.  

Specifically, we will assume that the full quantum dynamics has a spectrum of states corresponding to black holes, whose decay rates and spectra are approximately predicted by semiclassical Hawking radiation.  However, we do make one very important assumption that deviates strongly from this picture: we will assume that amplitudes for formation and evaporation of a black hole respect unitary evolution.  If this is correct, the thermal description of Hawking radiation must only be an approximation to more fundamental amplitudes where appropriate phases and correlations restore quantum purity. Then the black hole states should appear as resonances in scattering amplitudes.

 It appears that any dynamics that could produce such unitary evolution must have intrinsic nonlocality.  Indeed, \refs{\QBHB} has argued for a breakdown of perturbative gravity at an intermediate step in calculating Hawking's mixed-state density matrix\Hawkunc.  Whether or not there is some valid perturbative approach to calculating this density matrix, a plausible conjecture is that\refs{\LQGST,\GiddingsSJ,\GiddingsBE} {\it nonperturbative gravitational effects provide sufficient nonlocality to produce quantum-mechanical evolution, and unitarize amplitudes for black hole formation and evaporation.}

In the S-matrix context we do not need to understand {\it how} the dynamics respects unitarity, only {\it that} it does.  (In turn, as we'll discuss, properties of scattering respecting unitary evolution may furnish clues to the ``how" question.)  Among the quantum amplitudes representing different possible final states of our two-particle collision are the amplitudes for a two-particle final state, which will be of the form \tmat, \Tsimp, \pwsoln.  In general, properties of the phase shift $\delta_l$ and absorptive coefficients $\beta_l$ depend on properties of the intermediate states.  Thus in the regime of interest, $l\roughly<L(E)$, these will be related to properties of black holes.

\subsec{Black hole spectrum}

Basic features of the black hole spectrum appear to be the following.  First, a black hole of energy $E$ and momentum $l$ has a Bekenstein-Hawking entropy $S_{BH}(E,l)$, which for $l\ll E R_S(E)$, takes the form
\eqn\bhent{ S_{BH}(E,0) = {R_S(E)^{D-2}\Omega_{D-2}\over 4 G_D}\ ,}
with $\Omega_{D-2}$ given in \vols.  For more complete angular-momentum dependence see the appendix.  We expect this entropy to characterize the density of quantum black hole states.  Specifically, the number of black hole states in an energy range $(E,E+\delta E)$ is expected to be 
\eqn\nstates{{\cal N}(E,E+\delta E;l) = \rho(E,l)\delta E\ ,}
where the behavior of the density of states $\rho(E,l)$ is determined by the entropy,
\eqn\densstates{\rho(E,l) =  e^{S(E,l)}\ .}
Here $S(E,l)$ also incorporates possible subleading corrections to the Bekenstein-Hawking entropy.
These states are to be thought of  as resonances, as they have characteristic decay widths
\eqn\gammadef{\Gamma(E,l) \sim 1/R_S(E)\ ,}
given by the typical time to emit a Hawking quantum.   In particular, note that the spacing between states is exponentially small as compared to their widths.  The comparatively large widths lead to a smooth average density of states \densstates.

\subsec{Exclusive amplitudes}

We are now prepared to investigate properties of the partial-wave amplitudes for the two-particle final state.  Since these correspond to what is classically described as black hole formation, we expect that absorption dominates the elastic amplitude.  In particular, the semiclassical picture is that the two incoming particles form a black hole, which then evaporates.  The probability that this black hole will evaporate into precisely a two-particle final state is expected to be of order $\exp\{-S(E,l)\}$; we conjecture that this basic feature is not modified in the full unitary dynamics.\foot{For related discussion, see \ArkaniHamedKY.}  This immediately leads to an ansatz for the absorption parameters in the black hole regime:
\eqn\bhaabs{\beta_l = S(E,l)/4\quad,\quad l\roughly< L(E)\ .}

While several important features of the scattering appear to depend on the absorption parameters alone, it is also of interest to explore possible forms for the phase shifts.  Here, we recall that formation of a resonant state contributes $\pi$ to the phase shift.  Thus, from the collection of black hole states described above, we could conjecture that the gross features of the phase shift are captured by an expression of the form
\eqn\bhaphase{\delta_l(E) = \pi \int^E dE \rho(E,l) \sim \pi e^{S(E,l)}\ .}

At first sight, \bhaphase\ may seem puzzling in another context, which is the identification of the phase shift with the time delay, through the formula
\eqn\timed{\tau(E,l)\approx {d\delta_l(E)\over dE}\ .}
This produces a time delay for decay into the two-particle state that is exponentially large in the entropy, and seems at odds with the expected Hawking decay time of the black hole, which is of order $R_S S(E,l)$.  However, note that this actually fits with the statement above, that the partial width of the black hole into the two-particle final state is of order $\exp\{-S(E,l)\}$.  This in turn means that the corresponding decay time into such an atypical final state is indeed exponentially long, and thus supports the conjectured form \bhaphase.

Thus, we conjecture that gross features of high-energy scattering amplitudes in the regime of strong gravitational dynamics are summarized by the amplitudes of such a {\it black hole ansatz}, 
\eqn\bhamp{ T_{BHA}(s,t) = {i\over2} \psi_\lam s^{2-D/2}  \sum_{l=0}^L (l+\lam) \cnl(\cos\theta) \left[1- e^{-2\beta_l(E)+ 2i\delta_l(E)}\right]\ ,}
with $\beta_l$ and $\delta_l$ approximated by \bhaabs, \bhaphase.

\subsec{Transition to black hole regime}

It is also of interest to understand the transition between the eikonal regime and that of strong gravity; in particular, this is where perturbative gravitational calculations apparently fail.  At our current level of understanding we can only outline the transition at the level of the expected change in the partial-wave amplitudes.  Specifically, imagine fixing $l$ and increasing $E$ into the regime where $l\sim L$. As we see from \dbeik, below this energy threshold the leading eikonal phase shifts are approaching a value of order
$\delta_l \sim L\sim E^{D-2}$.
However, past the threshold one expects the behavior \bhaphase.  Thus, at the threshold we apparently find a very rapid variation of the phase.  Part of this variation may be captured by the eikonal contributions corresponding to the higher-order corrections to the classical metric. 

The absorptive coefficients are also typically expected to jump, but not as severely. Moreover, their change is related to model-dependent properties of the scattering, as outlined in section three.  For example, in the case of string theory, tidal string excitation becomes an important contribution to the asymptotic structure of the states as one approaches the black hole threshold.  (The arguments of \refs{\LQGST,\GGM} however strongly suggest that string excitation occurs on different time scales than horizon formation, and thus does not critically modify the picture that strong gravity becomes relevant.)  Estimated absorptive coefficients for soft graviton emission and tidal string excitation are given in \softg\ and \betats, and can be
compared to the expected absorption coefficients \bhaabs\ in the black hole regime, which for fixed $l$ have energy dependence
\eqn\betabha{\beta_l^{BHA}\propto E^{(D-2)/(D-3)}\ .}
Note that the soft graviton absorption coefficients at $l\simeq L$ are comparable in size, in accord with a semiclassical picture in which a non-zero fraction of the collision energy is emitted in gravitational radiation.

\newsec{Bounds, analyticity, and locality}

High-energy scattering behavior of a theory can convey important information about its structure, in particular through the asymptotics of amplitudes at high energies, which contain information about locality of the theory. As we have described, there are some reasons to believe that non-perturbative gravity is intrinsically nonlocal.  We can take a different tack on this locality question by asking whether scattering amplitudes that have basic properties expected of a gravitational theory have asymptotics corresponding to those of local theories, or not.

To explore contributions to scattering from the strong gravity region, let us examine the black hole ansatz \bhamp.  Note that the second term in brackets is exponentially suppressed in the energy.  Moreover, the sum over the first term can be performed explicitly (see formulas given in \CFV, below eq.~(2.5)).  The result is an expression of the form:
\eqn\bhapprox{T_{BHA}(s,t)= {i\over 2} \psi_\lam s^{2-D/2}\left[\left({L\over 2} + \lam \right) C_L^\lam(\cos\theta) + \lam (1+\cos\theta) C_{L-1}^{\lam+1}(\cos\theta)\right] + {\cal O}(e^{-S(E)/2})\ .}

Begin by considering the total cross section, given by the optical theorem in terms of the $\theta=0$ limit of this expression.  
Evaluating \bhapprox\ at $\theta=0$,  and using (A.3) and the asymptotic behavior of the gamma functions, we find that for large $E$ and thus $L$, 
\eqn\sigtot{\sigma_{T,BHA} \sim R_S(E)^{D-2} + {\cal O}(e^{-S/2})\ .}
This is as expected for formation of an object of size $R_S(E)$.

While the total cross section receives contributions from the other regimes as well, one can imagine isolating the contribution \sigtot\ from, for example, Coulomb scattering, by focusing on the absorptive part of the cross section.  (Strong absorption due to tidal/bremsstrahlung effects can however be an additional confounding factor.)  In any case, our expectations are that contributions of other regimes do not {\it reduce} the total cross section from the value \sigtot.

This is a first hint of nonlocality of the dynamics.  Local quantum field theories satisfy the Froissart bound\refs{\Froiss}, whose $D$-dimensional version\CFV\ states that the total cross section has a bound of the form
\eqn\Froissbd{\sigma_T\leq (R_0 \log E)^{D-2}\ .}
for some constant $R_0$.  However, violation of this bound does not conclusively imply nonlocality, as one of the assumptions in deriving the Froissart bound is the existence of a mass gap, which is not expected in gravitational dynamics.

Another important bound in local theories is the Cerulus-Martin bound\refs{\CeMa}, which is a {\it lower} bound on fixed angle scattering, of the form
\eqn\cemabd{T(s,\theta) \geq e^{-f(\theta) E \log E}\ .}
The fixed-angle asymptotics of \bhapprox\ are readily found from asymptotic expressions for the Gegenbauer polynomials\Szego.  At angles $\theta L\gg 1$, we find
\eqn\feasymp{\eqalign{T_{BHA}(s,t)\sim {i\psi_\lambda\over 2^\lam (\lam-1)!}& {s^{2-D/2} L^\lam\over \sin^{\lam+1}\theta }\Biggl\{ \sin\left[(L+\lam)\theta -{\lam\pi\over 2}\right]\cr& + \sin\left[(L+\lam+1)\theta -{\lam\pi\over 2}\right]+{\cal O}\left({1\over L\sin\theta}\right)\Biggr\}+ {\cal O}(e^{-S/2})\ .}}
This expression does not  violate the Cerulus-Martin bound.\foot{This appears in contradiction to expectations expressed in \ArkaniHamedKY.}  While amplitudes from the regime $l>L$ also make contributions, there is no reason to expect that these contributions lead to cancellations such that the total fixed-angle amplitude violates the lower bound \cemabd.

However, the behavior \feasymp\ is interesting from another perspective.  In particular, at large energies, it exhibits the feature of not being {\it polynomial bounded}.  Specifically, if we  use $L\sim ER_S(E)$ and continue to complex $E$, we find an exponentially growing amplitude.  Or, if we rewrite $\theta$ in terms of $t$ and $s$, and continue to positive $t$ (which involves going around the Coulomb pole of the full amplitude), we find behavior of the form
\eqn\expgrowth{T_{BHA}(s,t>0)\sim e^{2 R_S(E) \sqrt t}\ ,}
again violating polynomial bounds.  (Here we reasonably assume no cancellation from the ${\cal O}(e^{-S/2})$ terms.)

Polynomial boundedness is in fact a way that locality enters into the {\it assumptions} in proving the Froissart bound.  Thus we have tracked the behavior \sigtot\ to a more primitive source, which is independent of the issue of a gap.  Indeed, boundedness of amplitudes in the complex energy plane is essential for {\it causality}, as usual discussion of dispersion relations shows.  The basic idea is that if we have a system with a linear response, such that the response $r(t)$ to a signal $s(t)$ is of the form
\eqn\lineq{r(t) = \int_{-\infty}^\infty dt' S(t-t') s(t')\ ,}
then causality implies the condition that the Fourier transform of $S(t)$ be analytic and bounded in the upper half plane.  Exponential growth, such as
\eqn\expgr{S(E)\sim e^{\tau Im E}\ ,}
corresponds to acausal behavior with a time {\it advance} of size $\tau$.  

Translated to the scattering context, and  considering forward scattering, $\theta=0$, then the scattered wave should not traverse the scattering region faster than a corresponding wave traveling at infinity at the speed of light.
 This condition likewise implies polynomial boundedness of the forward scattering amplitude $T(s,0)$.  This is actually not in conflict with the above statements, as the asymptotic expression \feasymp\ is not valid at zero angle.  Indeed, 
 we found a polynomial-bounded expression in evaluating \sigtot.

Violation of a polynomial bound at non-zero angle does {\it not} necessarily imply violation of causality, if the scattering is finite in range.  The reason is that a finite range scatterer can shorten the path the incoming wave takes to become an
outgoing wave if $\theta\neq0$.  For an interaction of fixed range $R$, this corresponds to a relative time advance  $\approx 2R\sin(\theta/2) \approx 2 R |q|/E$.  Note that this nicely accords with the behavior \expgrowth, with the identification that the characteristic scattering range is the radius of the strong gravity region.  

Thus, the black hole ansatz and behavior of the form \feasymp, \expgrowth\ respect a macroscopic form of {\it causality}, which is good -- were they to do otherwise, one would suspect basic inconsistencies.  However, the lack of a non-forward polynomial bound corresponds to an interaction range that grows with energy.  This is not {\it local} behavior by traditional measures.

Notice that these conclusions rely only on very general features of the expected behavior of high-energy gravitational scattering.  Thus, either some of these features are not present, or this behavior supports the statement that nonperturbative gravity is not local in the standard sense.  Note also that these conclusions really only apparently depend on the expected form  \bhaabs\ of the absorption coefficients.  Due to the large absorption that these parametrize, the expressions \bhapprox-\sigtot\ and \feasymp-\expgrowth\ appear essentially insensitive to the real phase shifts $\delta_l$.  Indeed, note that the black hole ansatz is very similar to black disk scattering.  
In particular, the elastic cross section is 
\eqn\sigel{\sigma_{el}(s)\propto \sum_{l=0}^\infty (l+\lam)  \left[ 1- 2 e^{-2\beta_l(s)}  \cos(2\delta_l(s))+ e^{-4 \beta_l(s)}\right] \approx \sum_{l=0}^L(l+\lam) +\calo(e^{-S/2})\ ,}
and, in the same units, the total absorption cross section is 
\eqn\sigabs{\sigma_{abs}(s) \propto \sum_{l=0}^\infty(l+\lam) (1-e^{-4\beta_l}) \approx \sum_{l=0}^L(l+\lam) +\calo(e^{-S/2})\ .}
From these, $\sigma_{el}\approx \sigma_{abs}$.  The elastic contribution is that of diffraction scattering and should exhibit the appropriate diffractive peak.  

As a final comment, we note that due to the masslessness of the graviton, we don't expect all the usual analyticity properties of a gapped theory.  In particular, cuts in the $s$ or $t$ planes can extend all the way to their origins.  This feature raises a possible subtlety with crossing symmetry, an issue we leave for future examination.

\newsec{Conclusion}

If quantum gravity admits a state corresponding to the Minkowski vacuum, and excitations about this state, an S-matrix investigation of its theory appears appropriate.  We have outlined features of such an S-matrix, at super-planckian energies, beginning first with large impact parameters, and then with impact parameters that enter the strong-gravity regime.  At larger impact parameters, the Born or Coulomb amplitudes match onto the eikonal amplitudes, and the latter appear to yield a unitary extension of the former to smaller impact parameters.  As one approaches the regime $b\sim R_S(E)$, diagrams corresponding to exchange of gravitational tree amplitudes also become important, matching a classical picture.  As one reaches $b\sim R_S(E)$, the perturbation series apparently diverges.  However, 
approximate validity of the classical picture in this vicinity suggests the utility of  a semiclassical description of black hole formation and evaporation for inferring general properties of scattering in the strong gravity regime $b\roughly<R_S(E)$.  This leads to a proposal for features of the S-matrix in this regime, in the form of a {\it black hole ansatz}, given in section four.  We have in particular investigated asymptotic properties of this ansatz in section five, and found that appear not to respect a standard form of locality, namely  polynomial boundedness.  This serves as additional evidence for the lack of a fundamental role for locality in quantum gravity; such nonlocality of the nonperturbative theory may in turn be an important aspect of the internal description of black holes\refs{\GiddingsSJ,\GiddingsBE,\QBHB}, and may play an important role in cosmology\refs{\ArkaniHamedKY,\GiddingsNU}.  Such features of high-energy gravitational scattering may thus be supplying important clues towards a fundamental formulation of a quantum-gravitational theory.

\appendix{A}{Partial waves}

This appendix will provide further details of the partial-wave expansion.  We begin by making contact with the expansion for the transition matrix element given in \refs{\Sold}:
\eqn\Texp{T(s,t) = {\gamma_D s^{2-D/2}} \sum_{l=0}^\infty {1\over N_l^\lam} \cnl(1) \cnl(\cos \theta) f_l(s)\ ;}
recall $\gamma_D$ is defined in \gammaDef,
and $N_l^\lam$ is the normalization factor for the Gegenbauer polynomials,
%checked with mathworld
%
\eqn\nldef{ \int_{-1}^1 d(\cos\theta) (\sin\theta)^{D-4} \cnl(\cos \theta) C^\lam_{l'}(\cos \theta) = N_l^\lam \delta_{ll'}= {2^{1-2\lam} \pi \Gamma(l+2\lam)\over l! (\lam+l)\Gamma^2(\lam)} \delta_{ll'}\ .}
Using
\eqn\Cnone{
C_l^\lam(1) = {\Gamma(l+2\lam)\over \Gamma(l+1) \Gamma(2\lam)}\ ,}
\Texp\ can be simplified to the form given in the main text\CFV, eq.~\Tsimp.
The expression \pwderiv\ for the partial waves in terms of the amplitude can then be found using the orthogonality relation \nldef.  

It is also useful to have an expression for the delta function:
\eqn\deltaexp{\sum_{l=0}^\infty {1\over N_l^\lam} \cnl(1) \cnl(\cos\theta) = {2\delta(\cos\theta - 1)\over \sin^{D-4}\theta}\ .}
Here, accounting for the finite range of $\cos\theta$, we adopt the convention
\eqn\deltaconv{\int_0^1 dx \delta(x-1) = \hf\ .}

Finally, we record here the zero-argument value of the Gegenbauer polynomials:
\eqn\Czero{\eqalign{C_l^\lam(0) &= (-)^{l/2} {\Gamma\left({l\over 2} +\lam\right) \over \Gamma\left({l\over 2}+1\right) \Gamma(\lam)}\quad ,\quad l={\rm even}\ ;\cr
&=0\quad ,\quad l={\rm odd}\ .}}

\appendix{B}{Rotating black holes}

In this appendix, we collect some basic formulas for rotating $D$-dimensional black holes.   We will consider the case with only one non-zero rotation parameter.  We will not give the full form of the metric, which was first given in \refs{\MyersUN}.  
These solutions are have a characteristic ``Schwarzschild" radius $R_S(E,l)$, determined by solving the equation
\eqn\rotsch{R_S^{D-5}\left(R_S^2+{(D-2)^2l^2\over 4E^2}
\right) = 
{16 \pi G_D E \over (D-2) \Omega_{D-2}}\ . }
Subsequent formulas are simplfied by defining a rotation parameter,
\eqn\rotpar{a_* = {(D-2)l\over 2E R_S}\ .}
The Hawking temperature of the black hole is 
\eqn\thawk{T_H = {D-3 + (D-5)a_*^2 \over 4\pi R_S(1 + a_*^2) } \ }
and the Bekenstein-Hawking entropy referred to in the main text is 
\eqn\BHfull{S_{BH}(E,l) = {E\over (D-2) T_H}\left(D-3-{2a_*^2 \over  1 +a_*^2}\right)\ .}

\bigskip\bigskip\centerline{{\bf Acknowledgments}}\nobreak

We wish to thank N. Arkani-Hamed, T. Banks, K. Bardacki, D. Gross, A. Maharana, S. Mandelstam, and G. Veneziano for valuable discussions. 
The work of SBG was supported in part by the Department of Energy under Contract DE-FG02-91ER40618,  and by grant  RFPI-06-18 from the Foundational Questions Institute (fqxi.org).  The work of MS was supported in part by the National Science Foundation under grant PHY04-56556.

\listrefs
\end